# Application of the concept of hierarchical subordination to chain reactions in a nuclear reactor

V. V. Ryazanov

Institute for Nuclear Research, pr. Nauki, 47 Kiev, Ukraine, e-mail: vryazan19@gmail.com

Based on the theory of hierarchical structures, a correspondence has been established between the dynamics for the number of neutrons obtained from the theory of branching processes, the number of neutrons of the *n*-th generation, the number of nodes at the *n*-th level of the hierarchy, the rate of change in the probability of a chain reaction, the type of intensity and strength of the hierarchical connection, the degree reactor criticality, and neutron trajectories in the reactor. A connection has been found between the probabilities of the formation of a certain generation of the number of neutrons and the probability of the occurrence of a self-sustaining chain reaction of nuclear fission. It is shown that the Tsallis and Rényi distributions describing these processes are related by relations of deformed algebra, and under certain conditions can be accompanying with respect to each other.

## 1. Introduction

In [1], strict relations of the theory of percolation on Bethe lattices describe the behavior of the neutron multiplication factor. The critical point of the reactor corresponds to the percolation threshold. The behavior of the percolation probability, interpreted as the probability of the occurrence of a self-sustaining chain reaction, and its derivatives are considered. A striking manifestation of the complexity and nonequilibrium of chain nuclear processes in a reactor is their hierarchical structure. In this work, the statistics of hierarchical systems is applied to a more detailed study of complex fission chains.

Concepts of hierarchical subordination have been used to describe physical, biological, economic, environmental, social and other complex systems. One of the most productive applications of the idea of hierarchical structure is complex networks [5]. Real networks have a high degree of clustering and a self-similar structure, manifested in a power-law probability distribution over the number of connections between different neighbors [3, 6]. Many networks have a block structure, in the presence of which it is possible to identify groups of nodes that are strongly connected to each other, but have weak connections (or are completely unconnected) with nodes that do not belong to this group. This is due to the fact that the phase space of the system, far from equilibrium, when ergodicity is lost, is divided into clusters corresponding to structural levels that are hierarchically subordinate to each other. This is how fission chains behave in nuclear reactors. Hierarchically subordinate systems form an ultrametric space [2-4, 11]. Its geometric image is the Cayley tree (Fig. 1)

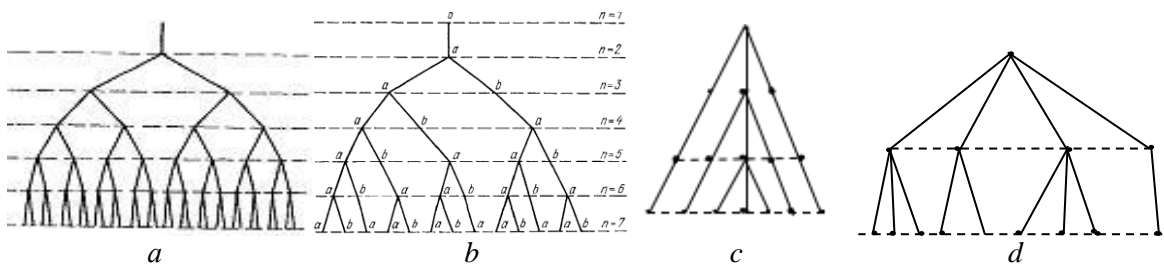

Fig.1. *a*). The simplest regular Cayley tree with branching s=2; b). Irregular Fibonacci tree with variable branching; c). Degenerate tree with s=3; d). Irregular tree for n=2, *a*=2.

In this work, some results of the theory of hierarchically subordinate systems far from equilibrium are applied to the description of nuclear chain reactions in a nuclear reactor. In the second section, a connection was discovered between the percolation features of the behavior of neutron-nuclear processes in reactors, considered in [1], with the intensity of the hierarchical object at level *n*, which for a stochastic system is reduced to probability density, and with the degree of hierarchical connection of objects *w*, corresponding to the nodes of the tree at a given level. For various hierarchical trees, an explicit form of these quantities is found, a correspondence is established with various operating modes of the reactor, with the trajectories of neutron motion in these modes. In the third section, the processes of anomalous diffusion in ultrametric space are considered, stationary solutions are found in the form of the Tsallis distribution [7], it is shown that these distributions in a particular case are escort in relation to the distributions found in the previous section (they are Rényi distributions) associated with probabilities of percolation and probabilities of a fission chain reaction.

## 2. Relationship between neutron reactor modes and neutron trajectories.

For neutron processes in a nuclear reactor, the most important characteristics are the percolation probability, which is interpreted as the probability of a self-sustaining chain reaction, and the percolation threshold value, which is proportional to the neutron multiplication factor. In [9], a recurrence relation was obtained for the probability of percolation from a root vertex, the probability that a connected component of the configuration containing the root vertex (some starting point of the appearance of the first neutron in the system, which generated a chain reaction), reaches the opposite edges of the system. Conventionally, mathematically, the size of the system and the connected component tends to infinity, although real systems are finite. In [9], the value $P(n,c)$ denotes the probability of percolation from the root vertex to a distance $n$. The value $n$ in our problem is interpreted as the number of generations of neutrons in a chain reaction. The number $c_\infty = \inf\{c: P(c)>0\}$ is called the percolation threshold in [9]. In [28], this value is called the critical probability at which a cluster first appears, extending over the entire lattice. Here $P(c)=\lim_{n\to\infty} P(n,c)$. In [1], for the percolation probability $P(n,c)$ a recurrence relation of the form was used:

$$P(n+1,c)=c[1-(1-P(n,c))^s], \qquad P(0,c)=c, \qquad (1)$$

where $c=p=\lambda_f(\lambda_f+\lambda_c)^{-1}$ is the probability of nuclear fission by a neutron, the intensity of neutron death (absorption by the environment or leaving the system) during the time $\Delta t \to 0$ is designated as $\lambda_c \Delta t + 0(\Delta t)$, and the intensity of nuclear fission by a neutron $\lambda_f \Delta t + 0(\Delta t)$ ($\lambda_f = v\Sigma_f$, $v$ —neutron speed, $\Sigma_f$—macroscopic fission cross section), $s = \overline{V}$, where $\overline{V}$ is the mathematical expectation of the number of secondary neutrons in one fission event. Effective neutron multiplication factor $k_{ef} = p\overline{V}$. The probability $c=p$ from (1) is associated with an important value of the percolation threshold, which is associated with the critical point of the reactor. Relations (1), as shown in [1], allow us to determine the critical point.

$$P_{n-1} = P_n + N_n^{-1} w(P_n), \qquad (2)$$

where $P_n$ is the intensity of a hierarchical object at level $n$, which for a stochastic system is reduced to probability density, this is the joint probability of the formation of an ensemble of hierarchical levels, an $n$-level hierarchical structure, $w$ is the degree of hierarchical connection of objects corresponding to tree nodes at a given level, $N_n$ is the number nodes at level $n$. The degree of hierarchical connection $w$ of objects corresponding to tree nodes at a given level is determined by the number of steps $n$ to the common ancestor, which specifies the distance in ultrametric space. The value $n$ in our case corresponds to the number of generations of neutrons in the fission chain reaction. The value $w$ corresponds to kinship in genealogy. Comparing expressions (1) and (2), we find that

$$w(P_n) = N_n[1 - P_n - (1 - P_n/c)^{1/s}], \quad s = \overline{V}. \qquad (3)$$

In [9] and [1], the value $P(n,c)$ denotes the probability of percolation from the root vertex to a distance $n$. In [1], this value is compared with the probability of a self-sustaining chain reaction. The quantity $n$, the number of generations of neutrons in a chain reaction, is proportional to time with a proportionality coefficient depending on the type of reactor. For thermal neutron reactors, the lifetime of one generation of neutrons is 0.1 sec; for fast neutron reactors, the lifetime of one generation of neutrons is 3-7 orders of magnitude less. The value $N_n$, - the number of nodes at level $n$, corresponding to the number of neutrons of the $n$-th generation, for a regular tree (Fig. 1a) is equal to

$$N_n = \overline{V}^n. \qquad (4)$$

The main feature of hierarchical systems is the property of self-similarity [8]. Let us consider the degree of hierarchical connection $w(P_n)$ (3) for small values of the argument. Expanding quantity (3) into a series in the region of small values $P_n \to 0$, for the value $P_{n0} \to 0$, $P_{n0} < P_n$, we find the maximum term of the expansion, which is equal to

$$w(P_n) = N_n A P_n^{1/s}, \tag{5}$$

where $A = P_{n0}^{(s-1)/s}(1 - P_{n0}/c)^{(1/s)-1}/c$, $P_{n0}$ is some fixed value of $P_n$, close to 0.

If we compare (5) with the expression obtained in [8] for the case $n \gg 1$, when $P_{n-1} \sim P_n$

$$w(P) = W P^\beta, \quad P \to 0, \tag{6}$$

where $W = w(1)$ is a positive constant, $\beta = 1-D$, $D \leq 1$ is the fractal dimension of a self-similar object such as an indented coastline [10, 11], we obtain that $1/s = \beta$, $D = 1 - 1/s = \ln\bar{V}/\ln q^{-1}$, $q < 1$ is the similarity parameter, and $P_n \sim q^n$, the connection function satisfies the homogeneity condition $w(qP) = q^\beta w(P)$. We find that at $\bar{V} = 2.4$, $\ln q^{-1} = \ln \bar{V}/(1 - 1/\bar{V}) \approx 1.5$, $q = (\bar{V})^{-\frac{1}{1-1/\bar{V}}}$. From a comparison of (5) and (6), since $W = w(1) = 1-c$, we also obtain that $P_{n_0} = \dfrac{[(1-c)c]^{\frac{\bar{v}}{\bar{v}-1}}}{(N_{n_0})^{\frac{\bar{v}}{\bar{v}-1}} + \dfrac{1}{c}[(1-c)c]^{\frac{\bar{v}}{\bar{v}-1}}}$. Assuming in equalities (2), (4) that for arbitrary values of $P_n$ the scaling relation $P_n = x_n q^n = x_n s^{-n/D}$ is satisfied, we arrive at the recurrent equality for the function $x_n$:

$$x_{n-1} = \Phi(x_n), \quad \Phi(x) = q(x + W x^{1-D}). \tag{7}$$

The mapping $\Phi(x)$ has two stationary points corresponding to the condition $x = \Phi(x)$: stable $x_s = 0$ and critical

$$x_c = (W/(q^{-1}-1))^{1/D}, \quad q = s^{-1/D}. \tag{8}$$

The behavior of the system is represented by homogeneous functions

$$P_n = x_c s^{-n/D}; \quad w_n = W^{1/D}(q^{-1}-1)^{-\Delta s - \Delta n}, \tag{9}$$

where $\Delta = (1-D)/D$ is the decrement that determines the scale of the hierarchical connection in ultrametric space [8, 11], which takes into account the vertices of hierarchical trees.

In [8], the continuum limit $n \to \infty$ is used, the finite difference $P_n - P_{n-1}$ is replaced by the derivative $dP_n/dn$, and an equation of the form (2) is written in continuous form. A comparison of the exact numerical calculation and solutions of approximate analytical expressions shows their convergence as $n$ increases, and coincidence already at values of $n$ of the order of 10-20. Let us consider solutions separately for different types of hierarchical trees with different behavior of the function $N_n$, the number of nodes at level $n$, corresponding to the number of neutrons of the nth generation. For small values of $P$, in asymptotics (6) for a regular tree with $N_n$ of the form (4), an explicit solution of this equation of the form

$$\begin{aligned} P &= W^{-1/(1-D)}[(1-u) + u e^{\zeta - \zeta_0}]^{1/D}, \\ u &= D W^{1/(1-D)}/\ln s, \\ \zeta &= (n_0 - n)\ln s, \\ \zeta_0 &= n_0 \ln s, \quad n \leq n_0, \\ w &= [(1-u) + u e^{\zeta - \zeta_0}]^\Delta, \\ \zeta &\leq \zeta_0, \quad w(\zeta_0) = 1, \end{aligned} \tag{10}$$

where $\zeta$ is the distance in ultrametric space, $n_0 \gg 1$ is the total number of hierarchical levels. The argument $c$ from (1) enters into (10), (13), (14) through $w$ from (3) and (5) and $W = w(1)$. For a given configuration of a hierarchical tree, an important role is played by the fractal dimension $D$, the value of which determines the strength of the hierarchical connection $w(\zeta)$. In nonstationary systems, the similarity parameter $q$ changes with time, and $D(q)$ also changes. For such complex systems as the system of multiplying neutrons in a reactor, the hierarchical connection is multifractal in nature [12]. An essential role is played by the spectrum of values of $q$, over which the coupling strength $w_q(\zeta)$ is distributed with density $\rho(q)$. The total value of the coupling force is determined by the equality $w(\zeta)=\int_{-\infty}^{\infty} w_q(\zeta)\rho(q)dq$. An expression of the form (10) with a variable value of the fractal dimension $D(q)$ is used as the kernel $w_q(\zeta)$. The behavior of this function for the reactor, obtained by calculation, is shown in Fig. 1 in [12] and Fig. 2b in [1]. The given relations determine only the asymptotic behavior of the hierarchical system in the limit $1<<\zeta\lesssim\zeta_0$. The resulting asymptotics represents the qualitative nature of the behavior of the hierarchical system. To obtain exact solutions, one must proceed from finite-difference equations of the form (1), (2), using numerical methods, as in [1]. The distribution over hierarchical levels was studied in [8] and is given below; it is shown that the stationary probability distribution takes the form of Tsallis. Note that when using a distribution containing the lifetime [13], it is possible to obtain more general distributions, in particular, superstatistics and their generalizations [14].

Tsallis distributions are only a special case of superstatistics and their generalizations. The probability of the formation of a self-similar network, in our case, the occurrence of a self-sustaining chain reaction of nuclear fission, increases monotonically with decreasing $n$, taking a maximum value at the upper level $n=0$ (initial neutron), corresponding to the entire system. The single initial neutron in the reactor has the maximum probability of causing a chain reaction, although its real possibilities for this are not yet so significant. The evolution of hierarchical structures is considered in [8] as a process of diffusion on randomly branching trees, the structure of which is determined by the heterogeneity parameter, which is a measure of their complexity. The complexity of a system, by analogy with entropy, characterizes the disorder of hierarchical communication [8]. But if entropy characterizes the disorder in the distribution of atoms, then when determining complexity, their role passes to sub-ensembles, into which the complete statistical ensemble is divided.

Relation (10) is written for the number of nodes $N_n$ at level $n$, corresponding to the number of neutrons of the $n$-th generation of type (4), $N_n=s^n$, where $s=\overline{V}$ is the branching index of the tree. We now use the above-mentioned proportionality of the number of generations of neutrons to time. Let us compare the expressions for the number of nodes $N_n$ at level $n$ with the temporal behavior of the number of neutrons, determined, for example, from the theory of branching processes [15, 16]. Expression (4) was written in [8] for the case of a regular tree shown in Fig. 1a and, since $n\sim t$, corresponds to the time behavior for the number of neutrons of the exponential form $e^{-\alpha t}$, valid outside the critical region [15]. It was shown in [16] that in the critical region the dependence is power-law, $t^\alpha$, which coincides with the power-law approximation of the form

$$N_n=(1+n)^a, \qquad (11)$$

used in [8] for the case of a self-similar irregular tree, Fig. 1d. This corresponds to a power-law dependence obtained in [1] numerically for the boundaries of the critical region. The behavior inherent in simple statistical systems is observed when the branching of the hierarchical tree exceeds the golden ratio $a_+=(5^{1/2}+1)/2\approx 1{,}61803$, and the decrease in complexity with increasing dispersion of the hierarchical connection, characteristic of complex systems, appears only with weak branching, limited to the interval $1< a <1.618$. For a degenerate tree, Fig. 1c,

$$N_n=1+(s-1)n\approx sn, \qquad (12)$$

which is close to (11) at $a=1$ and corresponds to a linear dependence on time and temporal behavior at the critical point [15, 16]. The question remains open whether movement along the Fibonacci tree corresponds to some physical situation in reactors, Fig. 1b [8]. In this case, fissile nuclei must be such that the average number of secondary neutrons produced during their fission is equal to the golden ratio $\tau=(5^{1/2}+1)/2\approx 1{,}61803$.

For a degenerate tree with the number of nodes (12), for behavior at the critical point, instead of the exponential dependence in (10), we obtain a logarithmic dependence of the form

$$P=W^{-1/(1-D)}[1-u\ln(1+(s-1)(\zeta_0-\zeta)/\ln s)]^{1/D},$$
$$u=DW^{1/(1-D)}/(s-1), \qquad (13)$$

$$\zeta=(n_0-n)lns, \quad \zeta_0=n_0 lns,$$

$$w=[1-uln(1+(s-1)(\zeta_0-\zeta)/lns)]^\Delta, \quad \zeta \leq \zeta_0.$$

In the intermediate case of an irregular tree with power-law growth (11) in the number of nodes (and neutrons), the intensity and strength of the hierarchical connection also behave in a power-law manner depending on the distance $\zeta$ in ultrametric space, proportional to the number of neutron generations:

$$P=W^{1/(1-D)}[1+u(1-\zeta/\zeta_0)^{-(a-1)}]^{1/D},$$

$$u=DW^{1/(1-D)}n_0^{-(a-1)}/(a-1),$$

$$\zeta=(n_0-n)lns, \quad \zeta_0=n_0 lns, \quad n \leq n_0, \qquad (14)$$

$$w=[1+u(1-\zeta/\zeta_0)^{-(a-1)}]^\Delta, \quad \zeta \leq \zeta_0.$$

The behavior of the probabilities of a chain reaction occurring is determined by the probabilities with, the degree of criticality, and the proximity to the critical point. Depending on this proximity, three (more precisely, four) main modes of behavior are distinguished: subcritical and supercritical (in them the laws of behavior (4) and (10) differ only in sign), critical (11), (14), and critical point (12 ), (13). In the traditional theory of nuclear reactors, only subcritical and supercritical regimes and the critical point are studied, although in the general theory of phase transitions a critical region is necessarily present. This is due to the fact that neutrons do not interact; the values of classical critical indices are valid for them (as for a self-consistent field) [12]. In the stationary operating state of reactors there are many neutrons, their number can be considered infinitely large. In this case, the critical region contracts to a critical point. Note that the explicit form of the expression $w(P_n)$ (2) is known, and the equation for $P$ in the continuum limit can be solved exactly. But the integrals are complex, and it is difficult to express the function $P$ explicitly.

The critical region itself has a complex three-member structure. In [16], three modes of critical behavior of nuclear reactors were discovered, depending on the sign of control actions and feedbacks, the boundaries of these modes were found, and it was shown that in the region of the critical point the time behavior is power-law. Time is proportional to the number of generations, and this behavior is characteristic of (11), self-similar irregular trees [8]. At the most critical point, the total number of neutrons is proportional to time (12), which corresponds to a degenerate tree. Thus, neutron trajectories vary depending on the probability $c$ and the multiplication factor. In the subcritical (and supercritical) region the movement occurs along regular trees, in the critical region along self-similar irregular trees, at the critical point along a degenerate tree. Above the critical point, but in the critical region - again using self-similar irregular trees. In the supercritical region - again using regular trees.

**3. Probabilities of formation of hierarchical levels, distribution by hierarchical levels and by neutron generations.**

Self-similar distributions are described by a power law of the form (11)

$$p(k) \propto k^{-\gamma} \qquad (15)$$

with exponent $\gamma > 0$, where $k$ is the degree of the tree vertex, which plays the role of scale in complex networks. Dependencies of this kind are widespread in systems of various natures. The form of such a dependence does not change with variation in the scale of the variable $k$, which determines the order distribution of the vertices of the hierarchical tree of a certain graph. Indeed, replacing the variable $k$ with a value $k/a$ scaled by a positive constant $a$ keeps the form of distribution (15) unchanged. Division chains lead to a hierarchical structure, the geometric image of which is the Cayley tree (Fig. 1).

In the general case, the cluster structure of all levels determines the behavior of the hierarchical system, but the self-similarity property allows us to limit ourselves to specifying the structure of the minimal cluster and finding the number of the hierarchical level. A hierarchical tree is a geometric image of ultrametric space [17], and in [3] it is shown that the description of hierarchical structures comes down to considering the process of diffusion in this space.

The evolution of complex hierarchical systems represents anomalous diffusion across hierarchical levels, which leads to a stationary distribution in the form of Tsallis (or Rényi distribution). Following [18], we consider the probability density of the distribution $p_u = p_u(t)$ of the system along the coordinates of ultrametric space at time $t$. This distribution obeys the kinetic equation [19, 20]

$$\tau \dot{p}_u = \sum_{u'} (f_{uu'} p_{u'} - f_{u'u} p_u), \qquad (16)$$

where the dot means differentiation with respect to time, $\tau$ is the relaxation time, $f_{uu'}$ represents the frequency of transitions from $u'$ to $u$. To determine the form of dependence on ultrametric coordinates, consider a regular hierarchical tree, which is characterized by a fixed branching index $s>1$ and a variable number of hierarchical levels $n>>1$. In this case, the ultrametric coordinate represents an $n$-digit number in the number system with base $s$: $u=u_0 u_1 \ldots u_m \ldots u_{n-1} u_n$, $u_m=0,1\ldots,s-1$. The intensity of transitions can be written in the form of a power series $f_{uu'} = \sum_{m=0}^{n} f(u_m - u_{m'}) s^{n-m}$, where the first term ($m=0$) corresponds to the upper level of the hierarchy, which determines the behavior of the entire system - the fission chain, while the last term with $m=n$ corresponds to the lowest level, corresponding to the smallest clusters, the last branches of the chain.

According to the definition, the distance between points $u$ and $u'$ is equal to $0 \leq l \leq n$ if the conditions $u_m = u_{m'}$ for $m=0,1,\ldots,n-(l+1)$ and $u_m \neq u_{m'}$ for $m=n-l, n-l+1,\ldots, n$ are met [11]. Thus, for a fixed distance $l$, the first $n-l$ terms of the indicated series are equal to zero by definition, while the last, the number of which is equal to $l$, contain the factor $s^{n-m}$, the value of which for $s>1$ is much less than the factor $s^l$, which is the first of the remaining terms. As a result, only the term with $m=n-l$ and $f_{uu'} \sim s^l = s^{n-m}$ is significant in the series under consideration. Similarly, it can be shown that the probability density is estimated as $p_u \sim s^{n-l} = s^m$. For a random tree, the branching index $s$ becomes variable, as a result of which the frequency of transitions $f_{uu'} \to f_{n-m}$ and the probability density $p_u \to p_m$ take the form of the Mellin transformation [20]

$$f_{n-m} \equiv \int_0^{\infty} f(s) s^{n-m} ds, \quad p_m \equiv \int_0^{\infty} p(s) s^m ds, \qquad (17)$$

where $f(s)$ and $p(s)$ represent weight functions. Thus, from the general coordinates $u=u_0 u_1 \ldots u_m \ldots u_{n-1} u_n$, $u_m=0,1\ldots,s-1$ of ultrametric space we move on to the coordinates of the level number, the number of neutron generations, which were used in the previous section.

As a result, the basic kinetic equation for the probability of formation of the $n$-th hierarchical level takes the form

$$\tau \dot{p}_n = \sum_{m>n} f_{m-n} p_n - \sum_{m<n} f_{n-m} p_m, \qquad (18)$$

where, in contrast to expression (16), which represents a continuous ultrametric space, a discrete representation is used that corresponds to hierarchical trees of the type shown in Fig. 1. The first term on the right side of (18) takes into account the hierarchical connection of a given level $n$ with lower levels $m>n$, the second with upper $m<n$. Noteworthy is the fact that the right side of equation (18) has the opposite sign to that in conventional statistical systems [21]. This is due to the fact that autonomous systems are characterized by the spontaneous establishment of a hierarchical connection, and not its destruction [3].

Expanding the probability $p_m$ in (18) into a series in powers of the difference $n-m$, in the limit $n>>1$ we obtain

$$\tau \dot{p}_n = -D(\partial^2 p_n / \partial n^2) + D_n p_n, \qquad (19)$$

where the lowest moments $\sum_{m<n}(n-m)f_{n-m}=0$ and $\sum_{m<n}(n-m)^2 f_{n-m} \equiv 2D$ are taken into account; the operator $D_n := \sum_{m>n} f_{m-n} - \sum_{m<n} f_{n-m}$ determines the difference in the intensities of transitions from a given level $n$ to lower and upper levels. If there is no hierarchy, then there are no conditions $m>n$, $m<n$ from (18), and the operator $D_n = 0$. In hierarchical systems, the intensity of transitions depends significantly on whether they occur up or down the hierarchical tree. We further use the assumption about the form of the function $D_n$.

$$D_n := -dq p_n^{q-1} \partial/\partial n, \qquad (20)$$

where $q$, $d$ are positive parameters. The formal basis of the assumption is that, up to a factor $-d(q-1)$, the integral $\int_n^{qn} D_n p_n \, dn$ reduces to the Jackson derivative

$$D_n p_n^q := \frac{p_{qn}^q - p_n^q}{q-1}, \qquad (21)$$

representing the archetype of self-similar hierarchical systems [4]. As a result, control equation (19) takes the final form

$$\tau \dot{p}_n = -(\partial/\partial n)\left(d p_n^q + D_n(\partial p_n/\partial n)\right). \qquad (22)$$

The stationary solution of this equation is written in the form of the Tsallis distribution [7]

$$p_n\left(p_0^{-(q-1)} + \frac{q-1}{\Delta} n\right)^{-1/q-1}; \quad p_0 \equiv \left(\frac{2-q}{\Delta}\right)^{\frac{1}{2-q}}, \quad \Delta \equiv D/d. \qquad (23)$$

According to (23), as the level number $n$ increases, the probability of its formation $p_n$ decreases in a power-law manner from the maximum value $p_0$ corresponding to the upper level $n=0$.

Using the deformed exponential $\exp_q(x) = [1+(1-q)x]_+^{1/1-q}$, $[y]_+ \equiv \max(y,0)$ and the effective energy $\varepsilon_n = \left(\frac{2-q}{\Delta}\right)^{\frac{q-1}{2-q}} n$, probability (23) takes the canonical Tsallis form

$$p_n = p_0 \exp_q\left(-\frac{\varepsilon_n}{\Delta}\right). \qquad (24)$$

According to [22], the effective temperature $\Delta$ satisfies standard thermodynamic relations, provided that the distribution over levels of a hierarchical self-similar set is determined by the escort probability $\mathcal{P}_l := \frac{p_l^q}{\sum_l p_l^q}$ and not the initial one $p_l$. It was noted in [23] and [8] that if we set $q`=1/q$, then the Tsallis escort distribution coincides with the Rényi distribution obtained by applying the maximum entropy principle to the Rényi entropy. In [24] it is shown that Renyi entropy serves as a negative indicator of the degree of conformal transformation of information discrepancy (divergence). The effective temperature $\Delta$ is related to the probability of nuclear fission from (1).

The probability of the formation of a hierarchical level and a self-similar chain (chain reaction) associated with this level increases monotonically with decreasing $n$. Calculation [8] demonstrates that the increase in dispersion $\Delta=D/d$, determined by the ratio of the diffusion coefficient $D$ to the energy $d$, significantly expands the spread of the stationary probability across hierarchical levels. For $\Delta\gg1$, distribution (23) differs slightly from exponential at high levels $n\ll\Delta^{1/2-q}$, but as $n$ increases, the power tail begins to manifest itself in an increasingly significant way.

Comparing distribution (23) with distribution (14), we see that one is an escort in relation to the other when $a=0$; these are the Tsallis and Rényi distributions. As already noted, these distributions will coincide when replacing the physical deformation parameter $Q=2-q$ with the value $2-q`$, $q`=1/q$. In this case, the dispersion $\Delta$ of the distribution (23) and the parameter $W$ in expression (14) are related by the dependence $\Delta=1/W$. The fractal dimension of ultrametric space in (14) is expressed through the deformation parameter $Q$:

$$D=q`-1=(Q-1)/(2-Q).$$

In the general case $a \neq 0$, and the indicated relations are of a particular nature. It is shown in [18] that the probabilities $P_n$ of an ensemble of hierarchical levels and the probabilities of the formation of each level $p_n$ are related by expressions of the so-called deformed (using the exponent $q$) algebra, when

$$\ln_q P_n = \sum_{m=0}^{n} \ln_q p_n,$$

$$\ln_q x = \frac{x^{1-q}-1}{1-q}, \quad P_n = p_0 \otimes_q p_q \otimes_q ... \otimes_q p_n,$$

$$x \otimes_q y = [x^{1-q} + y^{1-q} - 1]_+^{\frac{1}{1-q}},$$

$$P_n = \exp_q\left(\frac{\sum_{m=0}^{n} p_m^{1-q} - (n+1)}{1-q}\right) = (\sum_{m=0}^{n} p_m^{1-q} - n)_+^{\frac{1}{1-q}},$$

$$P_{n-1}^{1-q} - P_n^{1-q} = 1 - p_n^{1-q}.$$

The non-stationary case was considered in [8] in a self-similar mode, when the behavior of the system is determined by the power-law dependence $n_c(t)$ of the characteristic hierarchy scale (for example, the number of generations at which a percolation phase transition occurs, the critical point of the reactor), and the probability distribution is represented by a homogeneous function $p_n(t) = n_c^\alpha(t)\pi(n/n_c)$. Since in our case $n \sim t$, then depending on the type of $p_n(t)$ it is the self-similar regime that turns out to be significant.

### 3. Conclusion

The work presents a new approach to the study of complex processes in a nuclear reactor, based on synergetic methods associated with fractal and percolation methods for describing complex systems, and the theory of hierarchical subordination. New research methods make it possible to discover more detailed aspects of the behavior of reactor systems. Their comparison with traditional methods of studying neutron-nuclear processes in reactors will make it possible to find more subtle aspects of the behavior of these processes, take them into account and increase the safety of reactors.

Thus, the complexity of hierarchical trees in [25] was characterized by the silhouette $s_l=ln(M_l/M_{l-1})$, where $M_l$ is the number of nodes at level $l$. These expressions are given in (4), (11), (12). In a reactor, the ratio $M_l/M_{l-1}$ between the number of neutrons of neighboring generations characterizes the neutron multiplication factor. For regular trees (4) $s_l = ln\bar{v}$. At the most critical point $M_l=M_{l-1}$, and $s_l=0$, which corresponds to one neutron born in each generation and the picture of a degenerate hierarchical tree. For degenerate trees where $s=\bar{v}$

$$s_l = \ln[1 + \frac{(s-1)}{1+(s-1)(n-1)}] \approx \frac{(s-1)}{1+(s-1)(n-1)}.$$

This value tends to zero at $\bar{v}=1$ or at $n \to \infty$. For self-similar trees (11) $s_l=ln(1+1/l)^a \approx a/l$. This value tends to 0 as $l \to \infty$, which was noted in [1]. It is shown in [8] that a more adequate characteristic of the silhouette of a self-similar tree and the neutron multiplication coefficient corresponding to this value for a breeding reactor system is the Jackson derivative (21).

Let us also note that a powerful method for studying complex systems of this kind is the information geometry of probability distributions [26, 27]. The Rényi and Tsallis distributions discussed above are obtained by applying the maximum entropy principle to the Rényi $H^{(\alpha)}{}_R(p) = \frac{1}{1-\alpha} \log(\int p^\alpha(x) d\mu(x))$ species and Tsallis $H^{(\alpha)}{}_T(p) = \frac{1}{1-\alpha}(\int p^\alpha(x) d\mu(x) - 1)$ entropy. These entropies correspond to Renyi information deviations (divergences) $D^{(\alpha)}{}_R(p|q) = \frac{1}{\alpha-1} \log(\int q(\frac{p}{q})^\alpha d\mu(x))^{1/\alpha}$ and Tsallis type $D^{(\alpha)}{}_T(p|q) = \frac{1}{1-\alpha}(1 - \int p^\alpha q^{1-\alpha} d\mu(x))$ ($p$ and $q$ probability distributions). The generalized Pythagorean theorem [26, 27] is applicable to these quantities, to which expressions for the maximum entropy and other important physical results are attached, the application of which is essential for a detailed study of reactor systems.